\documentclass[pra,twocolumn,showpacs,superscriptaddress,nofootinbib]{revtex4-1}
\usepackage{graphicx}
\usepackage{bm}
\usepackage{amssymb}
\usepackage{color}
\usepackage[usenames,dvipsnames]{xcolor}
\usepackage{amsmath}
\usepackage{amstext}
\usepackage{latexsym}
\usepackage[colorlinks=true,citecolor=Cerulean,linkcolor=RubineRed,urlcolor=Cerulean]{hyperref}

\usepackage{amsfonts}
\usepackage{epsfig}
\usepackage{dsfont}
\usepackage{mathrsfs}
\usepackage{arydshln,leftidx,mathtools}
\begin{document}
\title{Exact Quantum Decay of an Interacting  Many-Particle  System: the Calogero-Sutherland model}

\author{Adolfo del Campo}
\address{Department of Physics, University of Massachusetts, Boston, MA 02125, USA}

\def\q{{\bf q}}

\def\G{\Gamma}
\def\L{\Lambda}
\def\la{\lambda}
\def\g{\gamma}
\def\al{\alpha}
\def\s{\sigma}
\def\e{\epsilon}
\def\k{\kappa}
\def\ve{\varepsilon}
\def\l{\left}
\def\r{\right}
\def\te{\mbox{e}}
\def\d{{\rm d}}
\def\t{{\rm t}}
\def\K{{\rm K}}
\def\N{{\rm N}}
\def\H{{\rm H}}
\def\la{\langle}
\def\ra{\rangle}
\def\om{\omega}
\def\Om{\Omega}
\def\vep{\varepsilon}
\def\wh{\widehat}
\def\tr{\rm{Tr}}
\def\da{\dagger}
\def\iz{\left}
\def\zi{\right}
\newcommand{\beq}{\begin{equation}}
\newcommand{\eeq}{\end{equation}}
\newcommand{\beqa}{\begin{eqnarray}}
\newcommand{\eeqa}{\end{eqnarray}}
\newcommand{\intf}{\int_{-\infty}^\infty}
\newcommand{\into}{\int_0^\infty}

\begin{abstract}

The exact quantum decay  of a one-dimensional Bose gas with inverse-square interactions is presented. The system  is equivalent to a gas of particles obeying generalized exclusion statistics. We consider the expansion dynamics of a cloud initially confined in a harmonic trap that is suddenly switched off. The decay is characterized by analyzing the fidelity between the initial and the time-evolving states, also known as the survival probability. It exhibits early on a quadratic  dependence on time that turns into a power-law decay, during the course of the evolution.  It is shown that the particle number and the strength of interactions determine the power-law exponent in the latter regime, as recently conjectured. 
The nonexponential character of the decay is linked to the many-particle reconstruction of the initial state from the decaying products.

\end{abstract}

\pacs{
03.65.-w, 
03.65.Ta, 
67.85.-d 
}
\maketitle
%

Understanding the decay dynamics of unstable isolated systems is of relevance to a wide variety 
of fields ranging from quantum science \cite{FGR78,Schulman08} to statistical mechanics \cite{Gorin06,EFG15,GE15} and cosmology \cite{KD08}. 
While the exponential decay law is ubiquitous in Nature \cite{time0}, quantum mechanics dictates its breakdown when the time of evolution is large \cite{Khalfin57} or small \cite{Ersak69}. The existence of these deviations follows from the linearity of unitary quantum dynamics. Consider the preparation of a unstable quantum state.
During its decay, the  time-evolving state  can be decomposed as a coherent superposition of the initial state and a set of decay products (any state orthogonal to the initial state).  Deviations from exponential decay result from the possibility for  the decay products to reconstruct  the initial state.

To be precise, let $|\Psi_0\ra=|\Psi(0)\ra$ be an unstable quantum state prepared at $t=0$, that evolves into $|\Psi(t)\ra$ when the dynamics is generated by a  Hamiltonian $H(t)$.
The probability to find the time-evolving state $|\Psi(t)\ra$ in its initial state $|\Psi_0\ra$ is referred to as the survival probability $\mathcal{S}(t)$, 
\beqa
\mathcal{S}(t):=|\mathcal{A}(t)|^2=|\la\Psi_0|\Psi(t)\ra|^2,
\eeqa
where $\mathcal{A}(t)=\la\Psi_0|\Psi(t)\ra$ is the survival amplitude. 
Equivalently, the survival probability is given by the expectation value of the projector on the intial state $\mathcal{P}=|\Psi_0\ra\la\Psi_0|$, 
\beqa
\mathcal{S}(t)={\rm tr}[|\Psi(t)\ra\la\Psi(t)|\mathcal{P}]
\eeqa
and is identical to the fidelity between  the initial state and the time-evolving state, $\mathcal{S}(t)\equiv F[|\Psi_0\ra\la\Psi_0|,|\Psi(t)\ra\la\Psi(t)|]$ \cite{Josza94}. 
The survival probability is as well closely related, but different, from the Loschmidt echo \cite{Gorin06}. However the distinction is often omitted in recent literature. 

To appreciate that the decay dynamics of $\mathcal{S}(t)$ is generally non exponential, it is convenient to use 
the Ersak equation, that relates the values of survival amplitude at different times during the course of evolution \cite{Ersak69}, 
 \beqa
 \label{eqErsak}
 \mathcal{A}(t)=\mathcal{A}(t-\tau)\mathcal{A}(\tau)+M(t,\tau).
 \eeqa
Above, the memory term is given by
  \beqa
 M(t,\tau)=\la \Psi_0|U(t,\tau)\mathcal{Q}U(\tau,0)|\Psi_0\ra,
 \eeqa
 where $U(t,t')={\bf T}\exp(-i\int_{t'}^tH(s)ds/\hbar)$ is the time evolution operator from time $t'$ to $t$, and $\mathcal{Q}=1-\mathcal{P}$ is the complement of the projector on the initial state $\mathcal{P}$. Hence, $M(t,\tau)$ represents the probability amplitude for the initial state to evolve into decay products at time $\tau$, and subsequently reconstruct the initial state at time $t$. For an exponential decay to hold, the memory term is to  vanish. 
However,  the memory term 
plays a dominant role at short and long times of evolution.
The short-time decay is known to be governed by the energy fluctuations of the unstable state. Experimentally, it was first demonstrated in \cite{shortexp} and its existence sets the ground for the quantum Zeno effect \cite{MS77}.
It is a consequence of unitary time-evolution, provided that the first and second moments of the Hamiltonian exist, see \cite{SPK12,CGC12} for exceptional cases.
That  deviations from exponential decay are as well to be expected at long times was pointed out by Khalfin in 1957, for systems whose  energy spectrum is bounded from below \cite{Khalfin57}. 
Measurements consistent with these deviations were reported in \cite{longexp}.

The decay dynamics of many-particle quantum systems has recently received a great deal of attention 
\cite{MG15,TS11,delcampo11,Goold11,GCML11,KB11,KLW11,PSD12,Longhi12, Hunn13,MK14}. These studies show the need to characterize quantum decay at a truly many-particle level, beyond its description in terms of one-body observables such as, e.g.,  the integrated density profile \cite{delcampo06,Zuern12,Rontani12,lattice}.
Achieving this is a challenging goal, due to the limitation of reliable numerical techniques.  Even at the single-particle or mean-field level, propagation methods based on space discretization in a finite spatial domain can introduce artifacts due to the enhanced reflection from the boundaries of the numerical box, unavoidable for long-time expansions \cite{Minguzzi04}.  An attempt to palliate this effect  with complex absorbing potentials \cite{complexV} explicitly suppresses state reconstruction, and delays the onset of power-law behavior in an unphysical way \cite{Muga06}. By contrast, these issues are absent in studies of fidelity decay in spin systems \cite{TS14}. As an outcome, analytical results in quantum decay, often based on time-dependent scattering theory, are highly desirable.
Recent theoretical progress has been mainly restricted to two particle systems \cite{MG15,TS11,GCML11,KB11,KLW11,Hunn13,MK14} and quasi-free few-particle quantum fluids  \cite{delcampo11,Goold11,Longhi12}.   Experimentally, the role of Pauli exclusion principle has been demonstrated using analogue simulation in photonic lattices \cite{Crespi15}, while interaction-induced particle correlations have been measured in optical lattices \cite{Preiss15}.

In this article, we present the exact quantum decay dynamics of an interacting many-body system  that is equivalent to a gas of particles obeying generalized exclusion statistics.
We rigorously show that the survival probability decays  as a power law  at long times with an exponent  that depends on the strength of the interactions and the particle number. In the non-interacting limit, this result proves the scaling conjectured based on the study of few particles systems \cite{TS11,delcampo11,GCML11,PSD12,MG15}. The non exponential character of the evolution is linked to the multi-particle reconstruction of the initial state.

\section{Model}
The Hamiltonian model we consider is that of $\N$ bosons effectively confined in one-dimension in a harmonic trap and interacting with each other through an inverse-square pairwise potential.  This is the so-called Calogero-Sutherland (CS) model \cite{Calogero71,Sutherland71}
\beqa
\label{HCS}
H(t)=\!
\sum_{i=1}^N\!\Big[-\frac{\hbar^2}{2m}\frac{\partial^2}{\partial q_i^2}+\frac{1}{2}m\om(t)^2q_i^2\Big]\!
+\sum_{i<j}\frac{\hbar^2\lambda(\lambda-1)}{m|q_i-q_j|^2}.
\nonumber
\eeqa
Its  energy spectrum (purely phononic)  and the complete set of eigenstates are known \cite{Calogero71,Sutherland71,VOK94}.
The CS Hamiltonian includes several relevant limiting cases. It reduces to  non-interacting bosons for $\lambda=0$, while the Tonks-Girardeau gas \cite{Girardeau60,GWT01,MG05,delcampo08} describing hard-core bosons  is recovered for $\lambda=1$. For $\lambda\neq\{0,1\}$, it describes interacting bosons which are equivalent to Haldane anyons \cite{Haldane91}.
We shall assume that the frequency of the trap for $t<0$ is given by $\om(t<0)=\om_0$. From now on, we use dimensionless variables $q\rightarrow q/q_0$, $t\rightarrow t/t_0$ with $q_0=(\hbar/m\om_0)^{1/2}$ and  $t_0=\om_0^{-1}$.

To describe the decay of the survival probability, we first note that the CS model belongs to a broad class of systems for which the exact time-dependent coherent states can be found \cite{Sutherland98,delcampo11b}. A  stationary state $\Psi$ of the system (\ref{HCS}) at $t=0$ with energy $E$,  follows a self-similar evolution dictated by the $SU(1,1)$ dynamical symmetry group,
\beqa
\label{scaling}
\Psi\left(q_1,\dots,q_\N,t\right)&=&\frac{1}{b^{\frac{\N}{2}}}\exp\left[i\frac{\dot{b}}{2b}\sum_{i=1}^\N q_i^2-i\frac{E}{\hbar}\tau(t)\right]\nonumber\\
& & \times \Psi\left(\frac{q_1}{b},\dots,\frac{q_1}{b},0\!\right)\!,\
\eeqa
where 
$\tau(t)=\int_{0}^tdt'/b^2(t')$. Here,  the scaling factor $b=b(t)>0$ is the  solution of the Ermakov differential equation
\beqa
\label{EPE}
\ddot{b}+K(t)b=b^{-3},
\eeqa
where $K(t)=[\om(t)/\om_0]^2$, and the boundary conditions $b(0)=1$ and $\dot{b}(0)=0$ follow from the stationarity of the initial state.
Note that $SU(1,1)$ and the scaling dynamics are robust against the breakdown of integrability \cite{Gambardella75,delcampo11b}.

The ground-state of the CS model has a Bijl-Jastrow form \cite{Sutherland71}
\beqa
\Psi(q_1,\dots,q_\N,0)=C_{\N,\lambda}^{-\frac{1}{2}}\prod_{i=1}^\N e^{-\frac{q_i^2}{2}}\prod_{j>i}|q_i-q_j|^{\lambda}.
\eeqa
Here,  the normalization constant $C_{\N,\lambda}$ reads
\beqa
C_{\N,\lambda}=
2^{-\frac{\N}{2}[1+\lambda(\N-1)]}
(2\pi)^{\frac{\N}{2}}
\prod_{j=0}^{\N-1}\frac{\Gamma\left(1+(j+1)\lambda\right)}{\Gamma\left(1+\lambda\right)};\nonumber
\eeqa
 it is related to the normalization constant of the probability distribution function for  the Gaussian $(\beta=2\lambda)$-ensembles in random matrix theory and 
derived using  Mehta's integral  \cite{Sutherland71,Mehta,Forrester10}.
Using the dynamics (\ref{scaling}), it is found that
\beqa
\mathcal{S}_{\N,\lambda}(t)&=&\frac{C_{\N,\lambda}^{-2}}{b^{\N[1+\lambda(\N-1)]}}\nonumber\\
& & \times \bigg|\int_{\mathbf{R}^{\N}}\prod_{i=1}^{\N}\! \d q_ie^{-\frac{q_i^2}{2}(1+\frac{1}{b^2}-i\frac{\dot{b}}{ b})}
\prod_{j>i}|q_i-q_j|^{2\lambda}\bigg|^2.\nonumber
\eeqa
Upon explicit computation, the following closed expression is obtained
\beqa
\label{SNanyons}
\mathcal{S}_{\N,\lambda}(t)=[\alpha(t)b(t)]^{-\N[1+\lambda(\N-1)]},
\eeqa
where the function $\alpha(t)$ is given by
\beqa
\alpha(t)=\frac{1}{2}\bigg[\left(1+\frac{1}{b^2}\right)^2+\left(\frac{\dot{b}}{b}\right)^2\bigg]^{\frac{1}{2}}.
\eeqa
As a result of the boundary conditions, $b(0)=1$ and $\dot{b}(0)=0$,
 $\alpha(t)$ reduces to unity at $t=0$. 
For $\N=1$, one recovers the survival probability of a single particle in a time-dependent harmonic trap $\mathcal{S}_{1,\lambda}(t)=\alpha(t)b(t)$.
It follows that the survival probability of a $\N$ particle CS system is identical to  that of
$\N$ non-interacting particles obeying Generalized Exclusion Statistics (GES) with exclusion parameter $g=\lambda$. 
GES was introduced by Haldane for systems with a finite Hilbert space \cite{Haldane91}, and extended by Wu to unbounded Hamiltonians \cite{Wu94}.
It accounts for the number of available states excluded by a particle in the presence of others, and it  smoothly extrapolates between bosonic and fermionic exclusion statistics, and beyond. Note that the many-body-wave function is always symmetric under the permutation of particles, this is,  the {\it exchange} statistics is always bosonic for arbitrary $\lambda$. The exclusion  parameter is defined as the ratio $g=-\Delta d/\Delta\N$ of the change in  the available states  $\Delta d$  as  the particle number is varied by $\Delta\N$.
For the CS model the exclusion parameter is precisely given by $\lambda$ 
\cite{MS94}. Particles with fractional excitation $\lambda\neq\{0,1\}$ are Haldane anyons.
From (\ref{SNanyons}), the survival probability for $N$ non-interacting and hard-core bosons is recovered for $\lambda=0,1$ respectively.
This leads to the duality relation
\beqa
\label{anyS}
\mathcal{S}_{\N,\lambda}(t)= [\mathcal{S}_{\N,0}(t)]^{1-\lambda} [\mathcal{S}_{\N,1}(t)]^{\lambda},
\eeqa
that represents a signature on the quantum decay dynamics imprinted by the transmutation of statistics observed in a CS system as a function  of the GES parameter $\lambda$.
Indeed, Eq. (\ref{anyS}) resembles the known relation for the {\it equilibrium} partition functions obeyed by Haldane anyons \cite{MS94}.
More generally,
\beqa
\label{anyS2}
\mathcal{S}_{\N,\gamma+\lambda}(t)= [\mathcal{S}_{\N,\gamma}(t)]^{1-\lambda} [\mathcal{S}_{\N,\gamma+1}(t)]^{\lambda}.
\eeqa
Simly put, these mathematical identities emphasize the fact that the CS model smoothly extrapolates between non-interacting and hard-core bosons (or more generally, between different types of Haldane anyons). In addition, this duality is not only apparent in equilibrium properties as discussed in \cite{MS94}, but can be clearly manifested in nonequilibrium observables as well.

\section{Sudden expansion}

\begin{figure}[t]
\begin{center}
\includegraphics[width=1\linewidth]{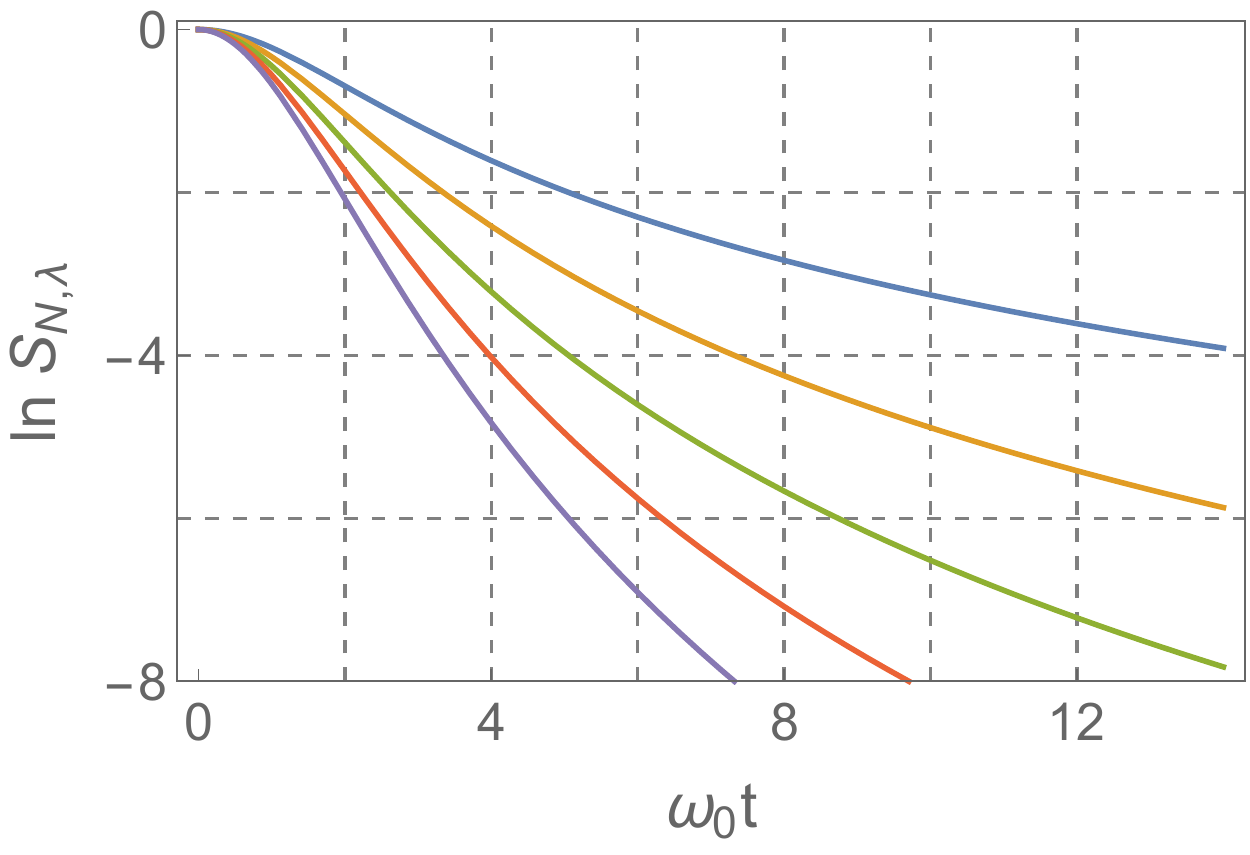}
\end{center}
\caption{\label{fig1}  {\bf Exact decay dynamics of a Calogero-Sutherland gas.}  The early stage exhibits a  quadratic dependence on  time and is followed by a smooth transition to the long-time power-law scaling. The dependence of the power-law exponent on the particle number ($\N=2$) changes from linear to quadratic as the interaction strength $\lambda$ is tuned from the non-interacting case to the Tonks-Girardeau limit ($\lambda=1$) and  other values $\lambda\neq \{0,1\}$, describing Haldane anyons with fractional exclusion statistics. From top to bottom, $\lambda=0,1/2,1,3/2,2$. 
}
\end{figure}

Suddenly switching off the trap (e.g., $K(t)=\Theta(-t)$), leads to an expansion dynamics with the scaling factor given by $b(t)=\sqrt{1+t^2}>0$.
The survival probability decays monotonically as a function of time and there is a smooth transition between the short and long time asymptotics,  as shown in  Fig. \ref{fig1}  for different values of $\lambda$. 
At short-times, $\mathcal{S}_{\N,\lambda}(t)$ is a quadratic function of time,
\beqa
\label{sasp}
\mathcal{S}_{\N,\lambda}(t)=1-\frac{1}{8}\N[1+\lambda(\N-1)]t^2+\mathcal{O}(t^4),
\eeqa
where all odd moments vanish identically.
Given the general short-time asymptotics of the survival probability
\beqa
\mathcal{S}_{\N,\lambda}(t)=1-\Delta H^2 t^2+\mathcal{O}(t^{3}),
\eeqa
where $\Delta H^2=\la\Psi_0|(H-E_0)^2|\Psi_0\ra$ with $E_0=\la\Psi_0|H|\Psi_0\ra$, the coefficient of $t^2$ in (\ref{sasp}) can be interpreted as the variance of the  energy in the initial state $|\Psi_0\ra$, $\Delta H=\sqrt{\N[1+\lambda(\N-1)]}/(2\sqrt{2})$.

As a result of the free-expansion dynamics, the exponential regime \cite{time0} is absent due to the lack of resonant states and a transition to 
the long-time behavior follows, when the survival probability is given by,
\beqa
\label{eqsaymp}
\mathcal{S}_{\N,\lambda}(t)&\sim&\left(\frac{2}{t}\right)^{\N[1+\lambda(\N-1)]}.
\eeqa
Hence, the survival probability decays as a power-law in time. The power law exponent depends linearly on the interaction strength $\lambda$ (the GES parameter) 
and exhibits an at most quadratic dependence on the particle number  $\N$.
This is the main result of this manuscript. Its derivation required (i) the scaling dynamics of the exact time-dependent coherent states, (ii) the use of the Bijl-Jastrow form of the initial state, and (iii) the identification of the leading term at long times.
The scaling dynamics holds exactly for the CS gas and simplifies the ensuing analysis by contrast to other many-body systems such as,  e.g., the 1D Bose gas, where only recently  moderate progress accounting for its dynamics has been reported \cite{Buljan08,Buljan11,IA12,Tetsuo12}.
Comparing the first leading terms in a long-time asymptotic expansion, it is found that
the power-law (\ref{eqsaymp}) sets in when the time of evolution satisfies
\beqa
\label{longt}
t\gg\sqrt{2\N(1+\lambda(\N-1))}.
\eeqa

%

For free-bosons, the power-law exponent becomes linear in the particle number
\beqa
\label{eqsaympB}
\mathcal{S}_{\N,0}(t)&\sim&(2/t)^{\N}.
\eeqa
The case of hard-core bosons correspond to $\lambda=1$, and leads to a power-law exponent quadratic in the particle number
\beqa
\label{eqsaympF}
\mathcal{S}_{\N,1}(t)&\sim&(2/t)^{\N^2}.
\eeqa
The change in the scaling with $\N$  was conjectured analyzing  quasi-free systems of $\N=2,3$ particles \cite{TS11,delcampo11,GCML11,PSD12,MG15}.  Eqs. (\ref{eqsaymp}), (\ref{eqsaympB}), and (\ref{eqsaympF}) prove that this is indeed the case.

\section{Robustness of the scaling}

It is worth emphasizing that the  power-law behavior (\ref{eqsaymp}) is observed in the long-time dynamics of other multi-particle observables such as the non-escape probability from a region of space, e.g., where the initial state is initially localized.
Explicitly, we define the $\N$-particle non-escape probability as
\beqa
\mathcal{P}_{\N,\lambda}(t)&:=&\int_{\Delta^{\N}}\prod_{n=1}^{\N}\! \d q_n|\Psi(q_{1},\dots,q_{\N};t)|^2,
\eeqa
which is the probability for the $\N$ particles to be found simultaneously in the $\Delta$-region and can be extracted from the full-counting statistics \cite{delcampo11}.
Explicit computation shows that
\beqa
\mathcal{P}_{\N,\lambda}(t)&=&\frac{C_{\N,\lambda}^{-1}}{t^{\N[1+\lambda(\N-1)]}}I_{\N,\lambda}(t).
\eeqa
Taking $\Delta=[-a/2,a/2]$,  the integral $I_{\N,\lambda}(t)$ becomes time-independent at long expansion times when $b\gg\sqrt{\lambda}a$, i.e. $t\gg|\lambda a^2-1|^{1/2}$,
\beqa
I_{\N,\lambda}(t)&:=&
\int_{\Delta^{\N}}\prod_{i=1}^{\N}\! \d q_ie^{-\frac{q_i^2}{b^2}}
\prod_{j>i}|q_i-q_j|^{2\lambda},\nonumber\\
&\sim&a^{\N[1+\lambda(\N-1)]}S_\N(1,1,\lambda),
\eeqa
where $S_n(\alpha,\beta,\gamma)$ is the Selberg integral \cite{Forrester10}. As  a result, the same power-law scaling sets in, i.e., 
\beqa
\mathcal{P}_{\N,\lambda}(t)\propto\mathcal{S}_{\N,\lambda}(t)\propto t^{-\N[1+\lambda(\N-1)]}.
\label{appsc}
\eeqa
However, the dependence of the power-law exponent on $\N$ and  $\lambda$ is lost when studying the decay in terms of one-body observables such as the  one-particle density profile $n(x,t)$ integrated over the region of interest $\Delta$,
\beqa
p(t):=\int_{\Delta}n(q,t) dq.
\eeqa
To illustrate this, let us consider the exact evolution of the density profile that under scaling dynamics is given by $n(q,t)=n(q/b,0)/b$.
Although an explicit computation of the density profile $n(q,0)$ is possible in the CS model, it would suffice to consider the large $\N$ limit. 
Then, $n(q,0)$  follows Wigner's semicircular distribution  $n(q,0)=\sqrt{1-q^2}2\N/\pi$ which is already independent of $\lambda$. Under free expansion,  it is found that $p(t)\sim 2 a\N/(\pi t)$, where the power-law exponent is independent of $\N$. The same conclusion holds when using the expressions for low $\N$ and large $\lambda$ available in the literature for $n(q,0)$ \cite{BF97,Forrester10}, but for the fact that the prefactor acquires a dependence on $\lambda/\N$. Generally, the $1/t$ power-law decay of $p(t)$ can be expected as the density profile flattens out at long expansion times, becoming approximately constant over the region $\Delta$, so that the integrated density profile $p(t)$ is governed by the normalization  factor $1/b(t)$.

One might also wonder whether a non-sudden  modulation of the trapping frequency will affect the long time power-law behavior. We show next that
as long as the frequency of the trap is permanently switched off after a given time $t=t_0$, the same power law scaling sets in.
Indeed, assume that $b(t_0)=b_0$, $\dot{b}(t_0)=v_0$, $K(t>t_0)=0$.  Then, 
\beqa
b(t)=\left[(b_0+v_0(t-t_0))^2+\frac{(t-t_0)^2}{b_0^2}\right]^{\frac{1}{2}},
\eeqa
 which tends to $b(t)\sim t(v_0^2+1/b_0^2)^{1/2}$ at large expansion times. As a result, only the prefactors of the survival and nonescape probability are affected, and the scaling is still dictated by (\ref{appsc}).

\section{ Many-particle state reconstruction}
%
\begin{figure}[t]
\begin{center}
\includegraphics[width=1\linewidth]{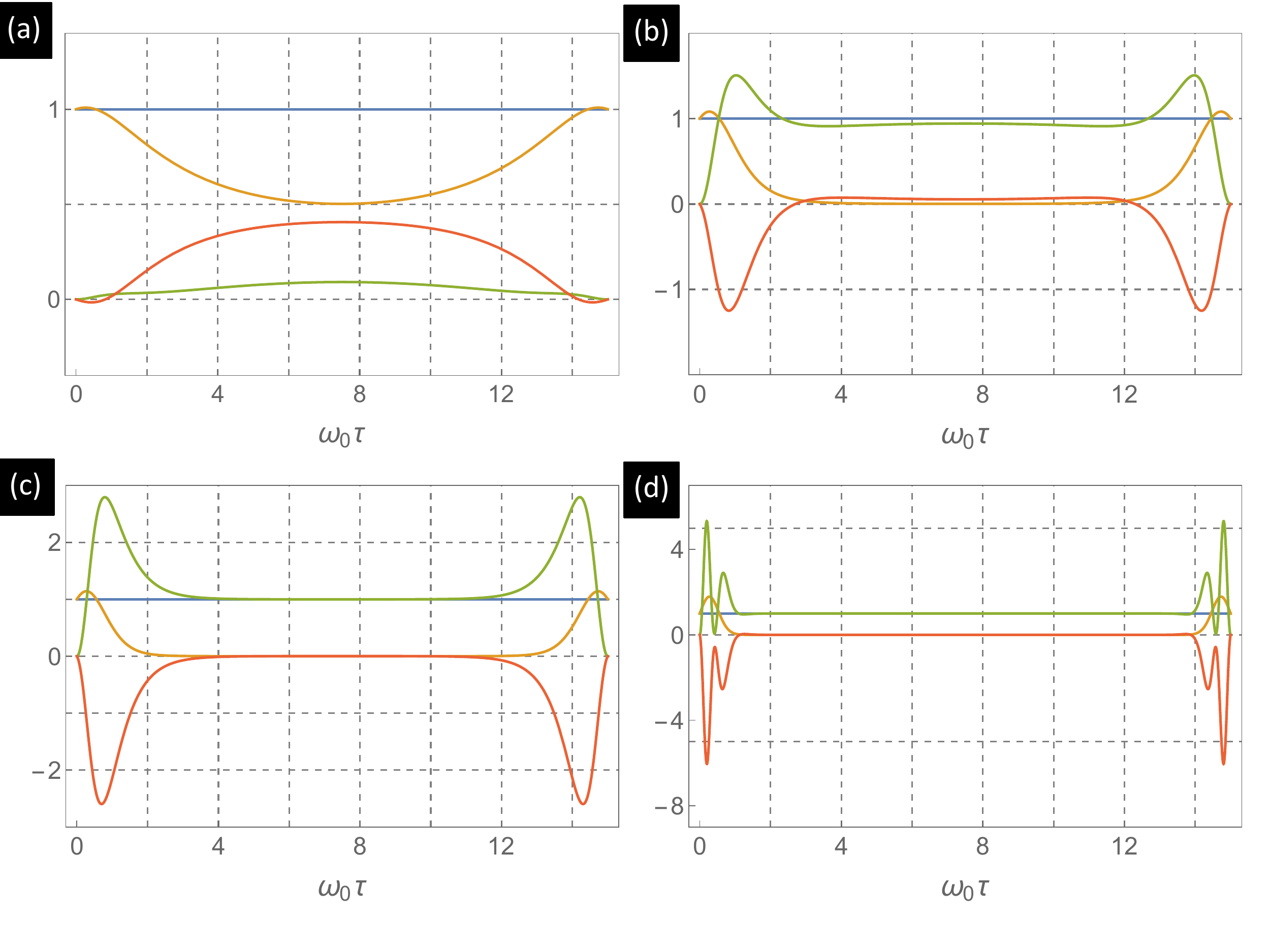}
\end{center}
\caption{\label{figsr}  {\bf Multi-particle state reconstruction.}  
The survival probability at a given evolution time is the sum of the product of the survival probabilities $\mathcal{S}_{\N,\lambda}(t-\tau)\mathcal{S}_{\N,\lambda}(\tau)$ (orange), the memory term $\mathcal{M}(t,\tau)$ (green)  and the interference term $\mathcal{I}(t,\tau)$ (red), as it follows from Eq.  (\ref{sqErsak}).  All contributions are normalized to normalized to the value at the final time $\mathcal{S}_{\N,\lambda}(t)$ and the blue line  is set at unity to identify the dominant term. 
(a) During the decay of a single particle all terms are significant.  (b) As the particle number is increased, the memory term becomes dominant ($\N=3$, $\lambda=1$). (c) For larger values of the interaction strength,  state reconstruction describes accurately the decay dynamics except for values of $\tau$ which are small or comparable to the total evolution time $t$ ($\N=3$, $\lambda=2$).  (d) Increasing the particle number ($\N=6$, $\lambda=2$) further prolongs the interval governed by state reconstruction. 
The dynamics is induced by suddenly switching off a harmonic trap with initial frequency $\om_0$, $t=15/\om_0$.
}
\end{figure}
We next analyze the relevance of state reconstruction in the CS model as an example of a many-particle system.
Using the Ersak equation (\ref{eqErsak}) \cite{Ersak69,Muga06}, the following decomposition of survival probability is obtained
\beqa
\label{sqErsak}
\mathcal{S}_{\N,\lambda}(t)=\mathcal{S}_{\N,\lambda}(t-\tau)\mathcal{S}_{\N,\lambda}(\tau)+\mathcal{M}(t,\tau)+\mathcal{I}(t,\tau),\nonumber
\eeqa
where  the first two terms admit a classical interpretation. In particular, $\mathcal{S}_{\N,\lambda}(t-\tau)\mathcal{S}_{\N,\lambda}(\tau)$ is the probability for the system to survive in the initial state at time $t$ provided that it was in the initial state at time $\tau$. Similarly, $\mathcal{M}(t,\tau)=|M(t,\tau)|^2$ accounts for state reconstruction in a classical sense, i.e., it is the probability that the state has decayed at time $\tau$ and reconstruct the initial state at time $t$. 
The last term in (\ref{sqErsak}) represents the interference between the amplitudes for the two histories just described, i.e., 
$\mathcal{I}_{\N,\lambda}(t,\tau)=2\Re[M(t,\tau)^*\mathcal{A}_{\N,\lambda}(t-\tau)\mathcal{A}_{\N,\lambda}(\tau)]$. 
Here, the survival amplitude is 
\beqa
\mathcal{A}_{\N,\lambda}(t)=\left[(b+\frac{1}{b}-i\dot{b})
\frac{e^{-i\tau(t)}}{2}\right]^{\frac{\N}{2}[1+\lambda(\N-1)]}.
\eeqa
Figure \ref{figsr} analyzes the relevance of each term in (\ref{sqErsak}) normalized to $\mathcal{S}_{\N,\lambda}(t)$ and after gauging away the dynamical phase $E_0\tau(t)/\hbar=N[1+\lambda(N-1)]\tau(t)/2$. Different regimes can be distinguished as a function of the parameter $\beta=\N[1+\lambda(\N-1)]$. For a fixed evolution time $t$  and small values of $\beta$ all terms in (\ref{sqErsak}) play a role. For larger values of  $\beta$, achievable by increasing either the particle number or the interaction strength,  $\mathcal{M}(t,\tau)$ dominates the contribution to the survival probability. Thus, the state reconstruction governs the long-time decay, except for values  of $t/\tau$ close to $\{0,1\}$, when all processes remain relevant. 

In conclusion, we have characterized the the exact  decay  of an interacting many-body quantum fluid released from a harmonic trap. Exploiting the self-similarity of the ensuing dynamics, the long-time power-law behavior of the survival probability was shown to be dictated  by the strength of the interactions even at arbitrarily large expansion times. The scaling of the power-law exponent is at most quadratic in the particle number. The non-exponential character of the decay can be attributed to the many-particle state reconstruction of the initial state.  Our results can be extended to other systems including  $SU(\nu)$ spin degrees of freedom 
and fermionic exchange statistics \cite{VOK94}.  As an outlook, it is worth  exploring  higher dimensional systems like the 2D Bose gas,
for which self-similar dynamics holds \cite{PR97} up to quantum anomalies \cite{OPL10} and the generalized exclusion parameter is known \cite{HLV01}.
In  systems lacking self-similar dynamics,  the role of interactions can be disentangled from that of the exclusion statistics and further studies will be illuminating.  A prominent example is the one-dimensional Bose gas with contact interactions, where recent advances in describing its dynamics have been reported \cite{Buljan08,Buljan11,IA12,Tetsuo12}.


{\it Acknowlegments.---} It is a pleasure to dedicate this article to Marvin D. Girardeau (1930-1915) and to thank 
S. B. Arnason, M. Beau, Y. Boretz, M. Cramer, T. Deguchi, V. Dunjko, K. Funo, J. Jaramillo, M. Olshanii, L. Santos and B. Sundaram for stimulating discussions.
Funding support from UMass Boston (project P20150000029279) is further acknowledged.


\end{document}